\documentclass[b5paper,twoside]{jpconf}  
\usepackage{graphicx}

\begin{document}

\title[An infrared telescope]{The First Infrared Telescope in Tibet Plateau, China}

\author[L.Y. Liu et al.]{Li-Yong Liu$^1$$^,$$^2$, Yong-Qiang Yao$^1$$^,$$^2$, Yi-Ping Wang$^1$$^,$$^2$, Jun-Rong Li$^1$$^,$$^2$, Yun-He Zhou$^1$$^,$$^2$, Lin Li$^1$$^,$$^2$, Xian-Long You$^1$$^,$$^2$}

\address{$^1$National Astronomical Observatories, Chinese Academy of Sciences, Beijing 100012, China}
\address{$^2$Key Laboratory of Optical Astronomy, Chinese Academy of Sciences, Beijing 100012, China}

\ead{liuly@nao.cas.cn}

\begin{abstract}

We plan to install an infrared telescope at the new site of Tibet,
China. The primary mirror diameter is 50$cm$, and the focal ratio
F8. The Xenics 640$\times$512 near infrared camera is employed,
equipped with a dedicated high speed InGaAs detector array, working
up to 1.7$\mu m$. The new site is located on 5100$m$ mountain, near
Gar town, Ali, where is an excellent site for both infrared and
submillimeter observations. The telescope will be remotely
controlled through internet. The goal of IRT is to make site
testing, detect variable stars, and search for extrasolar planets.

\end{abstract}

\section{The telescope and site}
The telescope made by Meade, a world leader in manufacturing of
amateur telescopes. Its primary mirror diameter is 50$cm$, and the
focal ratio is F8. A well sited 50$cm$ telescope could reach the 19
magnitude with a Deep Sky Imager(DSI) by 1 minute.

\begin{figure}[h]
\begin{minipage}{14pc}\hspace{1pc}
\includegraphics[width=13pc]{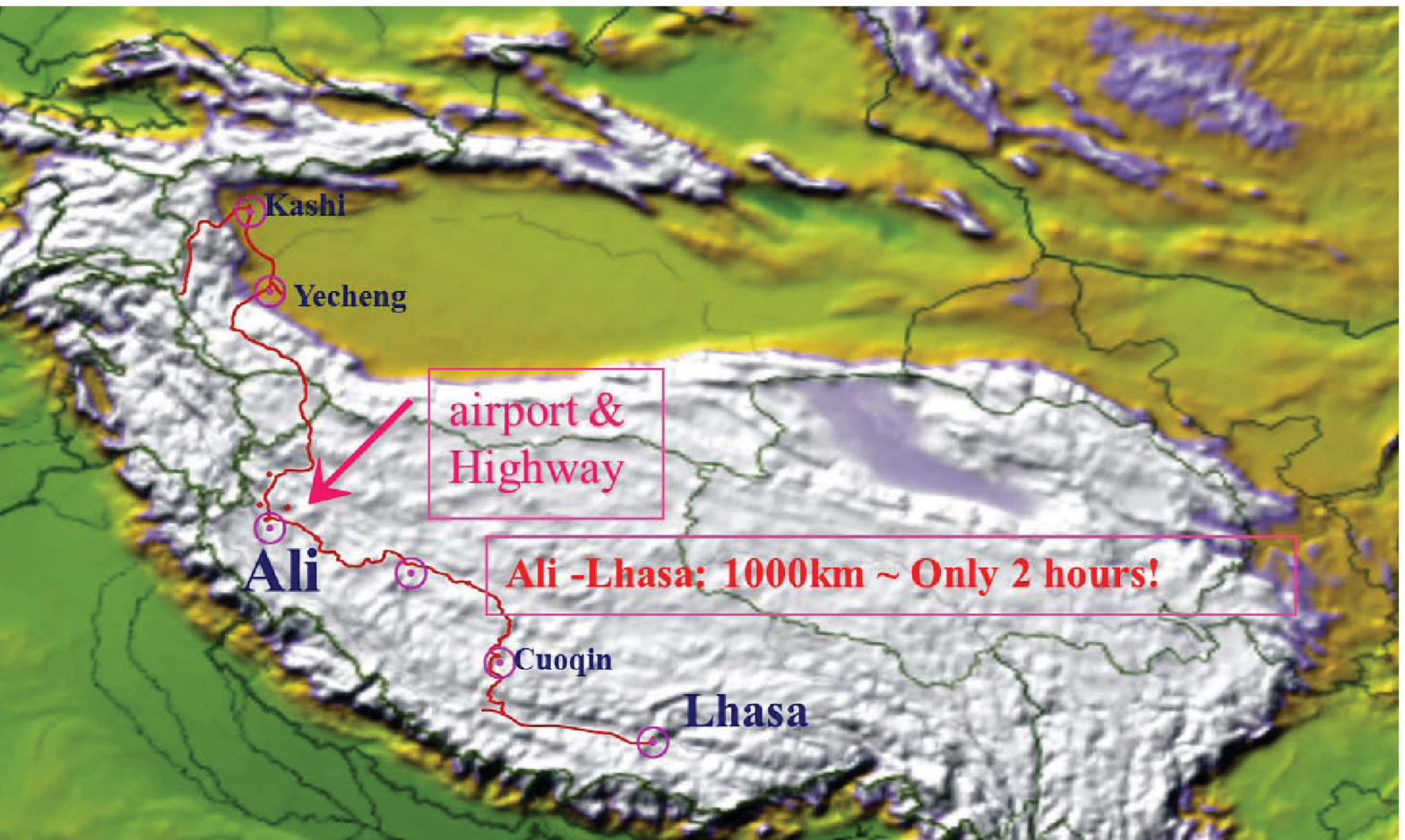}
\caption{The location of Ali site}
\end{minipage}\hspace{2pc}%
\begin{minipage}{16pc}\hspace{0.5pc}
\includegraphics[width=8pc,angle=270]{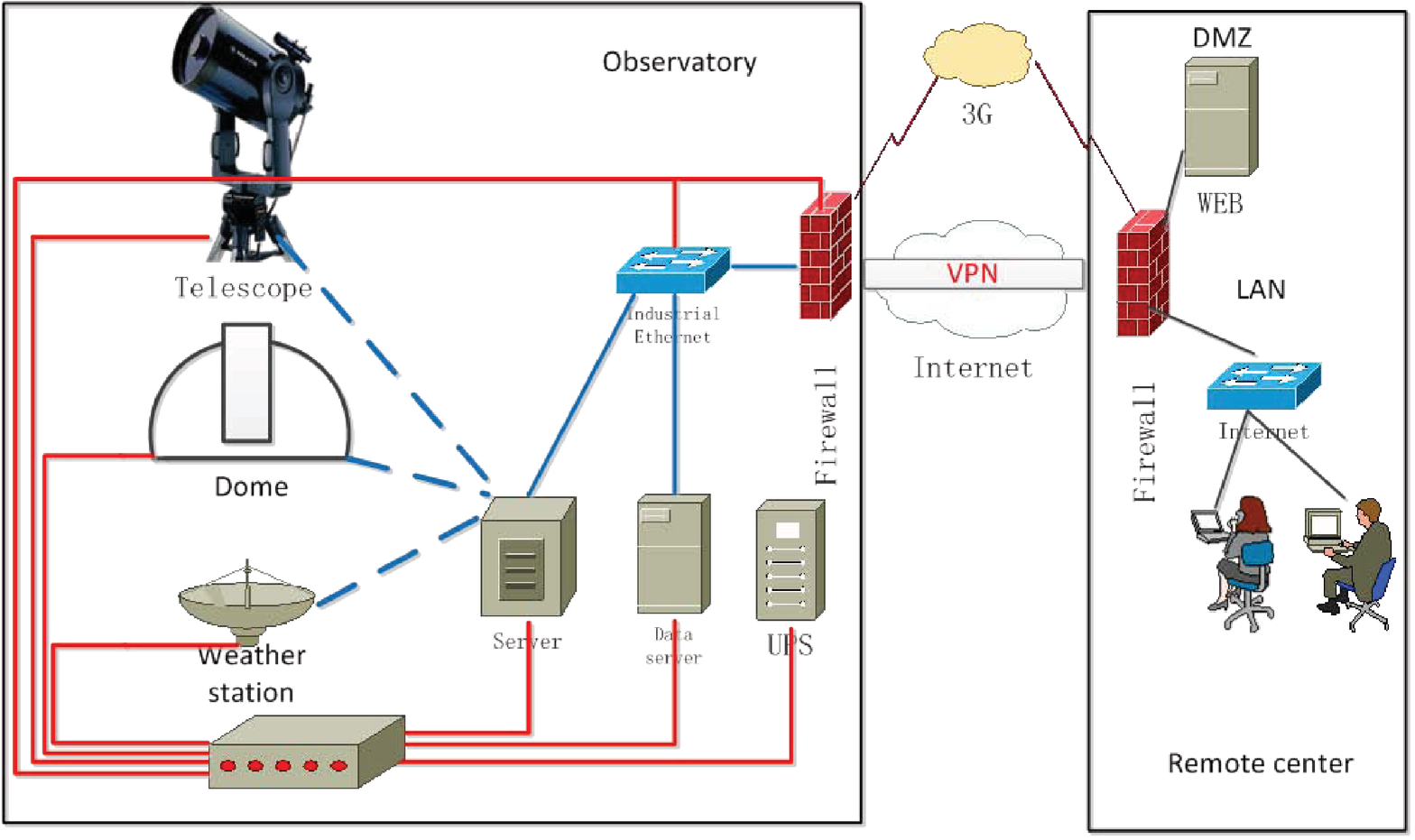}
\caption{The layout of remote control}
\end{minipage}
\end{figure}

The main instruments include a 4K$\times$4K optical camera, and a
640$\times$512 near infrared camera. The STX-16803 optical camera is
made by SBIG. The camera has a full frame image buffer for storing
image data during download. The XEVA-FPA-1.7-640 infrared camera is
one of the revolution in short wave IR cameras, working up to
1.7$\mu m$.

The candidate site selected is called Ali, Tibet(Fig1), located at
N32$^\circ$19$\verb|'|$, E80$^\circ$01$\verb|'|$, with altitude of
5100$m$\cite{yao08,liu08}. Remote study with the long-term database
of ground weather stations and archival satellite data has been
performed\cite{qian11,zhang10}. The site has enough relative height
on the plateau and is accessible by car.

\section{Remote and robotic operation}

The robotic telescopes are complex systems(Fig2) that combine a lot
of subsystems. These subsystems serve to provide telescope pointing
capability, control of telescope dome, as well as detection of
weather conditions. We are building a 6$m$ dome for the infrared
telescope. We have already installed satellite antenna in Beijing
and Ali for remote observations. The solar power system is also
equipped, and the introduction of high voltage power on the site is
possible.

\section{Goals of science}
\textbf{3.1 Site testing:} The telescope will be able to evaluate
Ali site of sky quality, atmospheric extinction, etc. It can also
get the characterizaiton of the astronomical seeing
conditions\cite{liu10}.
\\ \textbf{3.2 Experiment of high resolution target observations:} Fig3
is an image of the International Space Station, taken by a 20$cm$
telescope in Beijing. We hope to make further research of image
recovery with the 50$cm$ telescope on the excellent site.
\begin{figure}[h]
\includegraphics[width=17pc]{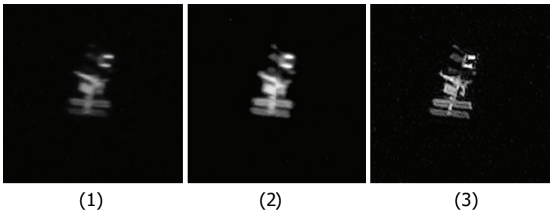}\hspace{0.8pc}%
\begin{minipage}[b]{14pc}\caption{\label{label}Fig3.1 is the original image from camera;
Fig3.2 is processed using Photoshop; Fig3.3 is the restoring object
images from the noisy turbulence degraded image of fig3.1.}
\end{minipage}
\end{figure}
\\ \textbf{3.3 Search for extrasolar planets:} We will use photometric
method to search for extrasolar planets, and we will also develop
some new methods to detect the atmospheric compositions of an
extrasolar planet.
\\ \textbf{3.4 Detection of variable stars:} Our goal is to detect
variable stars and classify these stars in an online catalog in
order to approach the origins of variations.

\section*{Acknowledgements}
This work was supported by the National Natural Science Foundation
of China (Grant Nos. 10903014, 11103042 and 11073031), and by the
Young Researcher Grant of National Astronomical Observatories,
Chinese Academy of Sciences.

\end{document}